\documentclass[aps,twocolumn,amsmath,amssymb,showpacs,prl,floatfix]{revtex4}
\usepackage{amsmath}
\usepackage{graphicx}
\usepackage{epsfig}
\newlength{\upit}\upit=0.1truein

\newcommand{\ltappr}{{{\lower4pt\hbox{$<$} } \atop \widetilde{ \ \ \ }}}
\newlength{\bxwidth}\bxwidth=1.5 truein

\begin{document}
\newcommand{\dg}{^{\dagger }}
\newcommand{\si}{\sigma}
\newcommand{\rarrow}{\rightarrow}
\def\fig#1#2{\includegraphics[height=#1]{#2}}
\def\figx#1#2{\includegraphics[width=#1]{#2}}
\newlength{\figwidth}
\figwidth=10cm
\newlength{\shift}
\shift=-0.2cm
\newcommand{\fg}[3]
{
\begin{figure}[ht]

\vspace*{-0cm}
\[
\includegraphics[width=\figwidth]{#1}
\]
\vskip -0.2cm
\caption{\label{#2}
\small#3
}
\end{figure}}
\newcommand{\fgb}[3]
{
\begin{figure}[b]
\vskip 0.0cm
\[
\includegraphics[width=\figwidth]{#1}
\]
\vskip -0.2cm
\caption{\label{#2}
\small#3
}
\end{figure}}

\newcommand \bea {\begin{eqnarray} }
\newcommand \eea {\end{eqnarray}}
\newcommand{\bk}{{\bf{k}}}
\newcommand{\bx}{{\bf{x}}}

\title{Singular conductance of a spin 1 quantum dot}

\author{A. Posazhennikova$^{1}$ and P. Coleman$^{2}$ }
\affiliation{$^{1}$ Institut  f\"ur Theorie der 
Kondensierten Materie, Universit\"at Karlsruhe, 76128 Karlsruhe, Germany \\
$^2$Center for Materials Theory,
Rutgers University, Piscataway, NJ 08855, U.S.A. }  
\pacs{72.15.Qm, 73.23.-b, 73.63.Kv, 75.20.Hr}
\begin{abstract}
We interpret the recent observation of a zero-bias anomaly in 
spin-1 quantum dots in terms of an  
underscreened Kondo effect.   Although a spin-1 quantum dots
are expected to undergo a two-stage quenching effect, in practice the
log normal distribution of Kondo temperatures  leads to 
a broad temperature region dominated by underscreened Kondo physics. 
General arguments,  based on the asymptotic decoupling between the
partially screened moment and the leads, predict a singular temperature and voltage
    dependence of the conductance $G$ and differential conductance $g$, resulting in
 $dg/dT\sim 1/T$ and $dG/dV \sim 1/V$. 
Using a Schwinger boson approach, we show how these qualitative
    expectations are borne out in a detailed many body calculation. 
\end{abstract}

%
\maketitle
%


Single-electron transistors (SETs) offer the intriguing opportunity to
probe and explore classes of strongly correlated electron behavior
associated with the Kondo effect that are
difficult to access in bulk materials\cite{physicsworld,raikh,lee,ggordon,koevenhoven}. The possibility of observing a
break-down in Landau Fermi liquid behavior that accompanies the {\sl
overscreened } two-channel Kondo effect in quantum
dots has been a subject of particular recent
interest\cite{twochannel1,twochannel2}.  
In this paper, we propose that singular deviations from Landau Fermi liquid
behavior associated with the {\sl underscreened} Kondo effect,
hitherto unobserved in bulk materials, will develop in conventional
quantum dots with even numbers of electrons and a triplet
ground-state\cite{schmid,sasaki,kogan}. These deviations from conventional Fermi liquid behavior
are predicted to lead to singular voltage, field and temperature
dependences in the conductance.


The Kondo effect in quantum dots with odd numbers
of electrons, predicted more than fifteen years
ago, \cite{raikh,lee}
is now well-established by experiment
\cite{ggordon,koevenhoven}. Subsequent
observations have shown that zero-bias
anomalies associated with a Kondo effect can also occur in quantum
dots with even occupancies, where Hund's coupling between the
electrons can lead to novel degeneracies, through the formation
of higher spin states, or the accidental degeneracy of singlet
and triplet states.  Zero-bias anomalies in 
integer spin  quantum dots were first reported by Schmid et al.\cite{schmid}.
Sasaki et al\cite{sasaki} later discovered a
zero-bias anomaly in even electron quantum dots, associated with the
degeneracy point between singlet and triplet states, tuned by a small magnetic field.
 Most recently,
Kogan et al\cite{kogan} have shown that the singlet-triplet
excitation energy in lateral quantum dots can be tuned by the gate
voltage, explicitly demonstrating that the zero bias anomaly develops
once the triplet state drops below the singlet configuration.

Pustilnik and Glazman\cite{pustilnik} have analyzed the low-temperature
Fermi liquid physics of higher spin quantum dots.  Their analysis
shows that lateral quantum dot in a triplet configuration
develops two screening channels which fully screen the local moment 
at the lowest temperatures. Using the Landauer formula, they deduce
the conductance $G$ of the Fermi liquid which develops to be 
\begin{equation}\label{}
G = \frac{2e^2}{h} \sin ^2 ( \delta_1 - \delta_2)
\end{equation}
where $\delta_1$ and $\delta_2$ are the scattering phase shifts of the two
screening channels. According to this line of reasoning, 
the development of a unitary phase shift in each channel, $\delta_1=\delta_2=
\pi/2$ leads to 
a complete suppression of the zero bias anomaly in a
triplet quantum dot\cite{zarand}. Why then, are 
zero-bias anomalies, with near unitary conductance  seen
in triplet quantum dots? 

In this paper we propose an interpretation of this unexpected behavior
in terms of an underscreened Kondo effect. Our key observation is that
the antiferromagnetic Kondo coupling constants $J_{\lambda}$
($\lambda=1,2$) associated with the two screening channels in a
triplet quantum dot will generally be distributed independently.
Since the Kondo temperature depends exponentially on the coupling
constant $T_{K\lambda}= D \sqrt{J_{\lambda} \rho } e^{-
\frac{1}{J_{\lambda}\rho }}$, a normal distribution of the coupling
constants will drive a log-normal distribution in the two Kondo
temperatures \cite{lognormal}, with the potential to generate
exponentially large separations in the relative magnitude of the Kondo
temperatures of each channel.  If we assume that $\ln
(T_{K1}/T_{K2})= \frac{1}{J_{2}\rho }- \frac{1}{J_{1}\rho }>>1$, then
over the {\sl exponentially broad} temperature range given by
$\log (T_{K1})>>\log T>>\log ( T_{K2})$, the underlying physics is
that of a one channel spin-1 Kondo model, in which the spin is
partially screened to a spin 1/2. 

From this perspective, triplet dots
with a large zero bias anomaly are those where the Kondo coupling
constants of the two channels are severely mis-matched, giving rise to
decades of behavior dominated by the under-screened Kondo effect in a
single channel.  Previous work, both
analytic\cite{pustilnik} and numerical \cite{hofstetter2,zarand} has
focussed on the equilibrium behavior of triplet quantum dots with 
Kondo temperatures of comparable magnitude.
We now examine the singular consequences of a wide 
separation between these two scales in both finite temperature and finite
voltage properties.

In the underscreened spin-1
Kondo effect, the residual spin-1/2 moment  is 
ferromagnetically coupled to leads, with a coupling that scales
logarithmically slowly to zero\cite{nozieres}.
The ground-state which develops is a ``singular Fermi
liquid'', in which the electrons do behave as Landau quasiparticles
which are elastically scattered with
unitary phase shift, but where, on the other hand, 
a logarithmically decaying coupling generates a singular
energy dependence in the scattering phase shift and a divergence in
the resulting quasiparticle density of
states\cite{pepin,pankaj,indranil}. 
The Hamiltonian for the underscreened quantum dot is 
\begin{eqnarray}\label{l}
H&=& H_{0}
+  J\psi \dg
_{\alpha}\vec{\sigma}_{\alpha \beta}\psi _{ \beta }
\cdot \vec{ S}, \nonumber \\ 
H_{0}&=&\sum_{k,\lambda= R,L ,\sigma}\epsilon_{k}
c\dg _{k\lambda\sigma}
c_{k\lambda\sigma} 
\end{eqnarray}
where $\psi_{\sigma } =\sum_{k }
\alpha c_{kL\sigma} + \beta c_{kR\sigma}
$
denotes the linear combination of right and left channels that couples
to the dominant screening channel. 
Much is known about the equilibrium physics of this model. 
At low temperatures, the spin is partially screened from spin $S$ to
spin $S- (1/2)$. The residual moment is ferromagnetically coupled to the
conduction sea, with a residual coupling that slowly flows to weak
coupling according to 
\begin{equation}\label{}
J\rho  (\Lambda) = -\frac{1}{\ln (
\frac{T_{K}}{\Lambda})
} +
O\left(\frac{1}{\ln^{2}(\frac{T_{K}}{\Lambda})
 } \right)
\end{equation}
where $\Lambda\sim max (T,\mu_{B}B)$ is the characteristic cut-off
energy scale, provided in equilibrium, by the temperature or magnetic
field. 
At low energies and temperatures, the partially
screened magnetic moment scatters electrons elastically, with a unitary phase
shift, however the coupling to the residual spin $( S- \frac{1}{2})$ gives rise
to a singular energy dependence of the scattering phase shift. The low
energy scattering phase shift can be directly deduced from the Bethe
Ansatz, and has the asymptotic form 

\begin{equation}\label{}
\delta(\omega) = \frac{\pi}{2}- \pi \rho J (\omega)= 
\frac{\pi}{2} \left(1 + \frac{( S-\frac{1}{2})}{\ln
(T_{K}/\omega)} \right).
\end{equation}

The logarithmic term on the right hand side is produced by the residual
coupling between the electrons and the partially screened moment. 
While the electrons at the Fermi energy  scatter elastically off the local moment with unitary
scattering phase shift, as in  a Fermi liquid, 
the logarithmically singular dependence of the phase shift leads to a divergent
density of states, $N (\omega)\sim \frac{1}{\pi}\frac{d \delta
(\omega)}{d\omega}\sim \frac{1}{\vert \omega\vert }$, which means that
we can not associate this state with a bona-fide Landau Fermi
liquid. For this reason, the ground-state of the underscreened Kondo
model has recently been called a ``singular Fermi
liquid''\cite{pankaj}. 

These singular features of the underscreened Kondo effect are expected 
to manifest themselves in the properties of the triplet quantum
dot. For example, we expect the low-field 
conductance to follow the simple relation
\begin{equation}\label{}
G (B) =  \frac{2e^{2}}{h}
\sin^2
\delta (B)\sim 
\frac{2e^{2}}{h}\left(1 - \frac{\pi^{2}}{16}
\frac{1}
{ln^{2} \left(\frac{T_{K}}{\tilde{B}} \right)} 
\right),
\end{equation}
for $ \tilde{B}<<T_{K}$.
This relationship was previously obtained by other means 
from the $T_{K2}\rightarrow 0$
of the two-channel model\cite{pustilnik}.
Notice that the field derivative of the conductance diverges as 
$\frac{dG}{dB}\propto 1/ \left(B ln^{3} 
({T_{K}}/{ \tilde{B}})
 \right)$
at low fields. 
The prediction of the finite temperature, and finite voltage
conductance can not be made exactly, however we expect the above form
to hold, for the differential conductance at finite temperature or
voltage, with an appropriate replacement of cut-offs, namely 
\begin{equation}\label{}
G (V,T) 
\sim 
\frac{e^{2}}{h}\left(1 - \frac{\pi^{2}}{16}
\frac{1}
{ln^{2} \left(\frac{T_{K}}{{\rm max} (T, eV)} \right)} 
\right)
\end{equation}
and 
$dG/dT\sim 1/ max (T,V)$. 

To  model this behavior in more detail it is useful to consider 
a simplified model of the quantum dot 
in which the 
Hund's coupling 
is taken to be infinite. In this limit, the states of the quantum dot
can be described using a Schwinger boson representation as 
\begin{gather}\label{l}
\vert d^{1}, \sigma \rangle = b\dg _\sigma\chi \dg \vert 0 \rangle ,
\qquad \qquad 
\vert d^{2}, M \rangle = 
 b\dg_\sigma b\dg_{M-\sigma}
\vert 0 \rangle , 
\end{gather}
Written in this representation the model becomes 
\begin{eqnarray}\label{}
H = H_{0} +  t \sum_{\sigma }\left[\psi \dg _\sigma \chi \dg b_{\sigma}
+b_{\sigma}\dg \chi\psi_{\sigma}
 \right] + E_{d}\chi \dg\chi 
\end{eqnarray}
subject to the constraint $n_{b}+\chi \dg \chi =2$. 

To develop a
controlled many body treatment of this Hamiltonian, we use a large-$N$
expansion, extending the number of spin components $\sigma $ from two
to $N$. 
To preserve a finite scattering phase shift as $N\rightarrow
\infty$, we introduce $K=kN$ bosonic ``replicas'', where $k$ is
fixed.  With this device we 
obtain a (dynamical) mean field theory with scattering phase
shift $\delta=\pi k$ and  the qualitatively correct logarithmic
energy dependences \cite{indranil}. 
The Hamiltonian used in the large
$N$ expansion is then
\begin{eqnarray}\label{}
H= H_{0}
+  \frac{\tilde{t}}{\sqrt{N}} \sum_{\sigma,\mu
}\left[\psi \dg _\sigma \chi_{\mu} \dg b_{\sigma}
+b_{\sigma}\dg \chi\psi_{\sigma}
 \right] + E_{d}\sum_{\mu=1}^{kN}\chi_{\mu} \dg\chi_{\mu}.
\end{eqnarray}

In the large N limit, there are 
two self-consistent non-crossing approximations to the Dyson equations
for the self-energies of the conduction electrons and $\chi $ fermions (Fig. 1). Here we sketch the main elements of the derivation. 
As in the corresponding equilibrium calculation\cite{indranil}, 
the boson behaves as a sharp excitation in the large $N$
limit, with an average occupancy $\langle n_{b\sigma \mu}\rangle = n_{b}/N$. 
From the Dyson equations we obtain sets of self-consistent integral
equations for both the retarded and Keldysh self-energies. 
The explicit expressions for the 
the retarded self-energies $\Sigma^{\chi}_{R}$ and $\Sigma^c_{R}$ of
the slave fermion 
($\chi$) and the conduction electrons ($c$) are 
\begin{widetext}
\begin{eqnarray}\label{l}
\Sigma^{\chi}_{R} (\omega) &=& - {\tilde t}^2 n_b G_A(\lambda-\omega)
 -{\tilde t}^2 \int \frac{d\omega'}{\pi}
f_c(\omega')\frac{1}{\omega'+\omega-\lambda+i\delta} ImG_R(\omega'), \cr
\Sigma^{c}_{R} (\omega) &=& - {\tilde t}^2 k n_b J_A(\lambda-\omega) -
{\tilde t}^2 k \int \frac{d\omega'}{\pi}
f_{\chi}(\omega')\frac{1}{\omega'+\omega-\lambda+i\delta} ImJ_R(\omega').
\end{eqnarray}
\end{widetext}
Here $J_R(\omega)=[\omega-E_d-\Sigma^{\chi}_R(\omega)]^{-1}$ and
$G_R^c(\omega)=[(-\pi \rho)^{-1}-\Sigma_R^c(\omega)]^{-1}$ are the
retarded propagators for the $\chi$ fermions and conduction electrons.

\begin{figure}[!tbh]
\begin{center}
\includegraphics[width=0.47\textwidth]{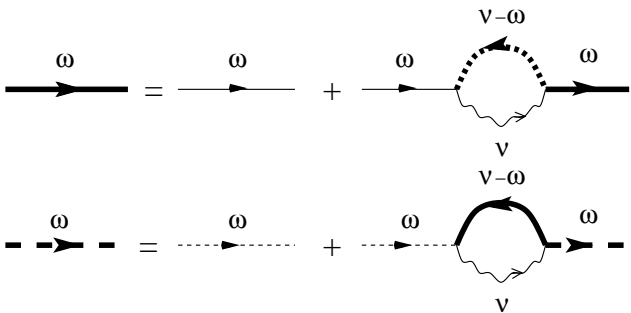}
\end{center}
\caption{The non-crossing approximation for the self-energies of conduction electrons 
and $\chi$ fermions. The solid line denotes the  Larkin Ovchinnikov
matrix propagator for the conduction electrons. The 
dashed line denotes the corresponding Green's function of 
the auxiliary ($\chi $) fermions and the wavy line is the
bosonic propagator. Thin lines denote the bare propagator and full
lines the dressed propagator. Each vertex corresponds to the factor $\frac{\tilde t}{\sqrt{N}}$.}
\label{NCA}
\end{figure}

The ratio of the Keldysh to the retarded self-energies
self-consistently determines the fermion distribution functions.
We can summarize the results of our calculation of the Keldysh self-energies
by providing the distribution functions that they generate. 
The distribution function of the conduction electrons is the average 
\begin{equation}
f_c=\frac{1}{2}\left[f_{L}(\omega)+f_{R}(\omega)\right],
\end{equation}
where $f_{L,R}(\omega)=1/(e^{\beta (\omega \mp eV/2} +1)$ is the equilibrium
distribution function in the left/right-hand lead.
The distribution function of the auxiliary fermion is 
\begin{equation}
f_{\chi} (\omega) =\frac{n_b[1-f_c (\omega)]}{n_b+f_c (\omega)},
\end{equation}
where $n_{b}= 1/ (e^{\beta \lambda}+1)$ determines $\lambda$. 
This relationship can be simply understood as the result of detailed
balance between rate of the decay processes $c\rightarrow b+\chi $ and 
$b+\chi \rightarrow c$, and it reverts to the equilibrium Fermi Dirac
distribution in the limit $V\rightarrow 0$.

\begin{figure}[!tbh]
\begin{center}
\includegraphics[width=0.5\textwidth]{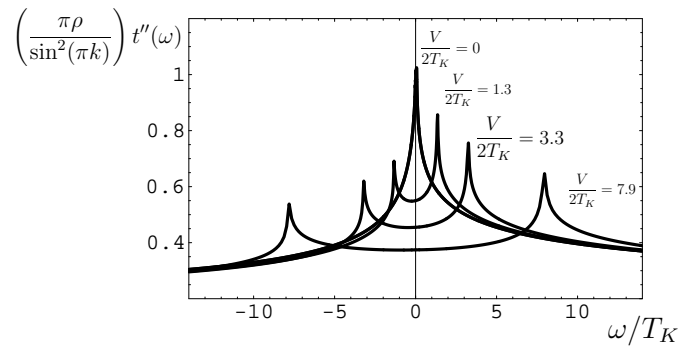}
\end{center}
\caption{
Imaginary part of dot T-matrix for a variety of voltages for the case $k=0.4$. As the
voltage is increased, the singular central peak splits into two
components. 
}
\label{fig2}
\end{figure}

From these results, we compute the temperature and
voltage dependent current, given by \cite{wingreenmeir}
\begin{equation}\label{}
I (V,T)= N \frac{e^{2}V}{h} \rho  \int {d\omega} \frac{[f_{L}
(\omega)-f_{R} (\omega) ]}{eV}
{\rm  Im} t_{R} (\omega)
\end{equation}
where  $t_{R} (\omega)=
\Sigma^{c}_{R} (\omega)/[1 - i \pi \rho \Sigma^{c}_{R} (\omega)]$ is
the scattering t-matrix.

We have solved these equations numerically, and the key results are
shown in Figs 2-4. Fig. 2. shows the voltage dependent t-matrix at
zero temperature. At zero voltage, the t-matrix contains a logarithmic
singularity noted in previous work\cite{hofstetter2,indranil} 
which splits into two peaks at a finite voltage.  In our calculation,
the split Kondo resonance retains its singular structure, although
this is most likely an artifact of taking a limit where the bosons
behave as a sharp excitation. In Fig. 3., we show the temperature
dependent conductance. 
\begin{figure}[!htb]
\begin{center}
\includegraphics[width=0.5\textwidth]{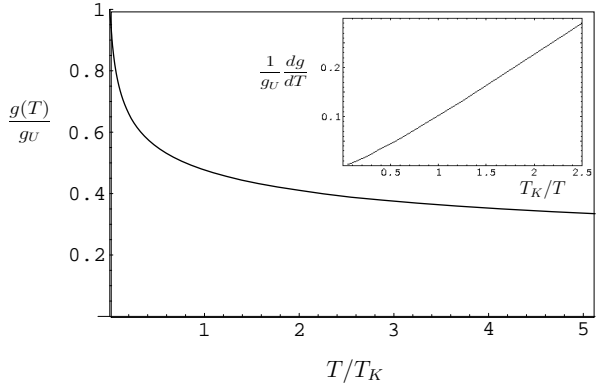}
\end{center}
\caption{
Temperature dependence of the differential conductance, normalized with respect to $g_{U}=N
\frac{e^{2}}{h}sin^{2} (\pi k)$ for the representative case $k=0.4$. 
Insert shows the $1/T$ divergence of the derivative $dg/dT$. 
}
\label{fig3}
\end{figure}
The temperature-dependent deviations from
unitary  conductance are determined by the logarithmic singularity
in the phase shift, and in our calculation, these are proportional to
$1/\ln  (T_{K }/{\rm max} (eV,T))$.  In the Schwinger boson approach, the number of
bound bosons in the Kondo singlet never exceeds $N/2$ and the region
$K\geq N/2$ does not describe an underscreened Kondo
model. Consequently, we are limited to static phase shifts 
 $\delta = \pi (K/N)< \pi/2 $, so the strictly particle-hole symmetric case $\delta =\pi/2$ is
outside the limits of our approach. Nevertheless,  our numerical
results do capture the expected singularities. Fig. 3. shows the
singular form of the temperature dependent conductance, with singular $1/T$
divergence in $dg/dT$. 
\begin{figure}[!htb]
\begin{center}
\includegraphics[width=0.45\textwidth]{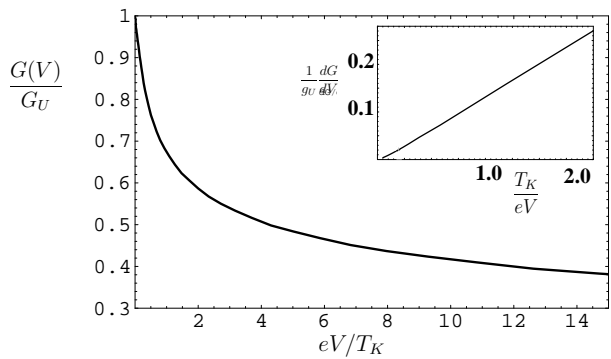}
\end{center}
\caption{
Voltage dependent conductance $G (V)=I (V)/V$ for the case
$k=0.4$. Insert: $dG (V)/dV$ showing $1/V$ divergence. 
}
\label{fig4}
\end{figure}
Finally, Fig. 4. shows the voltage dependence of the conductance,
which has a similar logarithmic singularity at low voltage.

In summary, we have proposed that the monotonically increasing
conductance observed as the temperature is lowered in triplet quantum dots
is associated with an underscreened Kondo effect.  The singular
energy and temperature dependence  associated with the Kondo resonance
is predicted to give rise to a $1/T$ divergence in the temperature
dependence of the differential conductance, and a $1/V$ divergence in the
second derivative of the voltage dependent current $d^{2}I/dV^{2}$.
These ideas have been developed qualitatively and illustrated within an
integral equation treatment of the underscreened Kondo model. Experimental observation
of these singular features would constitute a first 
realization of the underscreened Kondo effect. 


The authors wish to thank H. Kroha, G. Zarand and M. Eschrig for
discussions related to this work. This research was partly supported
by the Alexander Von Humboldt foundation (AP) and DOE grant
DE-FG02-00ER45790 (PC).


\end{document}